\documentclass{llncs}
\usepackage{amsmath}
\usepackage{psfrag}
\usepackage{graphicx}
\usepackage[ruled,linesnumbered,noend]{algorithm2e}

\title{Optimal Data Placement on Networks With Constant Number of Clients}
\author{Eric Angel\inst{1} \and 
        Evripidis Bampis\inst{1} \and 
        Gerasimos G. Pollatos\thanks{This work has been supported in part by the project BIONETS (BIOlogically inspired NETwork and Services) (FP6-027748)}\inst{2} \and 
        Vassilis Zissimopoulos\inst{2}}

\institute{Universit\'{e} d'\'{E}vry-Val d'Essone, IBISC CNRS FRE 2873, 523 place des Terrasses,
           91000 \'{E}vry, France\\
           \email{\{angel,bampis\}@ibisc.univ-evry.fr} \and
           Department of Informatics and Telecommunications, University of Athens, Greece\\
           \email{\{gpol,vassilis\}@di.uoa.gr}}

\newcommand{\fs}[1]{\footnotesize{#1}}
\newcommand{\comment}[1]{}

\begin{document}
\bibliographystyle{splncs}

\maketitle

\begin{abstract}
We introduce optimal algorithms for the problems of data placement (DP) and page placement (PP) in
networks with a constant number of clients each of which has limited storage
availability and issues requests for data objects. The objective for both problems is to efficiently
utilize each client's storage (deciding where to place replicas of objects) so
that the total incurred access and installation cost over all clients is minimized. In the PP
problem an extra constraint on the maximum number of clients served by a single client must be
satisfied. Our algorithms solve both problems optimally when all objects have uniform lengths. When
objects lengths are non-uniform we also find the optimal solution, albeit a small, asymptotically
tight violation of each client's storage size by $\varepsilon l_{max}$ where $l_{max}$ is the
maximum length of the objects and $\varepsilon$ some arbitrarily small positive constant. 
We make no assumption on the underlying topology of the network (metric, ultrametric etc.), thus
obtaining the first non-trivial results for non-metric data placement problems. 
\end{abstract}

\section{Introduction}
Peer-to-peer file sharing networks have become one of the most popular aspects of everyday internet 
usage. Users from all around the globe interact in an asynchronous manner, benefiting from the 
availability of the desired content in neighboring or more distant locations. The success of such 
systems stems from the exploitation of a new resource, different from the traditional bandwidth - 
related resources, namely the distributed storage. Widespread utilization of this new resource is 
due to the fact that larger capacities have become cheaper, with significantly smaller data access 
times. Interacting users, utilize this resource by installing local storage, replicating popular 
content and then making it available to neighboring users, thus dramatically decreasing bandwidth 
consumption, needed to access content from the origin servers at which it is available.

A suitable abstract model describing perfectly the aforementioned situation is the \textit{data
placement} problem (DP)~\cite{Baev01}. Under this model, a set of clients (equivalently
users or machines) with an underlying topology is considered and each client has a local amount of
storage (cache) installed. Given the set of available objects and the preference that each client
has for each object, the objective is to decide a replication scheme, also referred to as a
{\em placement} of objects to local caches so as to minimize the total access cost among all
clients and objects. The generalization of this model, under which each client's cache has an upper
bound on the number of clients it can serve is known as the \textit{page placement} problem
(PP)~\cite{Meyerson01}. 

It should be noted, that the term replication is used here instead of caching, because under the
discussed model, a client cannot change the contents of its local storage without reinvocation of a
replication algorithm. On the contrary, the term caching refers to the process of choosing objects
to store locally so as to serve requests and use a replacement scheme so as to replace some of them
for others on-the-fly according to popularity or other criteria. 

\vspace{8pt}

\noindent \textbf{Our contributions.} We descibe optimal algorithms, combining configurations 
generation and dynamic programming techniques, for the data placement and page placement problems,
when the number of clients is constant. This is a natural variation, interesting from both a
theoretical and practical point of view (\cite{Feldman99,Feldman99bis,Kimelfeld06}). Up to now, the
only way to tackle these problems, was the 10-approximation algorithm of~\cite{Baev08} and
the 13-approximation algorithm of~\cite{Guha02}), both designed for the general case and both based on 
rounding the solution of an appropriate linear program. When object
lengths are uniform (or equivalently unit) our algorithm finds the optimum solution in polynomial
time. When object lengths are non-uniform, our algorithm returns an optimum solution which violates
the capacities of the clients' caches by a small, asymptotically tight additive factor. 
Our results, summarized in table~\ref{tab1}, can be
modified to handle various extensions of the basic problems such as the connected data placement problem (\cite{Baev08}) where object
updates are frequent and consistency of all replicas of each object has to be guaranteed and
the $k$-median variant of DP where bounds are imposed on the number of maximum replicas allowed for each
object. Furthermore, our results are applicable with uniform and non-uniform object lengths and can be employed 
independently of the underlying topology of the network, thus giving the first non-trivial
results for non-metric DP problems.
\begin{table}[t]
\begin{center}
\begin{tabular}{|c||c|c||c|c|}
\hline
 & \multicolumn{2}{c||}{\textbf{\fs{Known results}}} & \multicolumn{2}{|c|}{\textbf{\fs{In this paper}}}\\ 
\hline
 & \multicolumn{2}{c||}{\textbf{\fs{arbitrary}} $M$} & \multicolumn{2}{|c|}{\textbf{\fs{fixed}} $M$} \\ 
\hline
 & \textbf{\fs{metric}} & \textbf{\fs{no metric}} & \textbf{\fs{metric}} & \textbf{\fs{no metric}} \\
\hline
\fs{uniform lengths DP}     & \fs{10-approx \cite{Baev01,Baev08,Guha02}} & - & \multicolumn{2}{c|}{\fs{optimal}}  \\
\hline
\fs{non-uniform lengths DP} & \fs{10-approx with}                        & - & \multicolumn{2}{c|}{\fs{optimal with}}  \\
                            & \fs{blow-up $l_{\max}$ \cite{Baev08}}      &   & \multicolumn{2}{c|}{\fs{blow-up $\varepsilon l_{max}$}}\\
\hline
\fs{page placement}         & \fs{13-approx \cite{Guha02}\,$^*$}         & - & \multicolumn{2}{c|}{\fs{optimal with}} \\
                            &                                            &   & \multicolumn{2}{c|}{\fs{blow-up$^{**}$ $\varepsilon
l_{max}$}}\\
\hline
\fs{connected DP}           & \fs{14-approx \cite{Baev08}}               & - & \multicolumn{2}{c|}{\fs{optimal}} \\
\hline
\fs{$k$-median DP}          & \fs{10-approx \cite{Baev01,Baev08}}        & - & \multicolumn{2}{c|}{\fs{optimal}} \\
\hline
\end{tabular}
\caption{\label{tab1}\fs{The main known results on data placement problems ($^*$non-uniform lengths
with constant blow-up on clients and cache capacities, $^{**}$non-uniform lengths
with constant blow-up on cache capacities only).}}
\end{center}
\end{table}

\noindent \textbf{Related work}. The study for the data placement problem over an 
arbitrary network where all inter-client distances form a metric was
initiated in~\cite{Baev01} where the authors proved that the problem in the case of objects of uniform
length is MAXSNP-hard. They also devised a polynomial 20.5-approximation algorithm based on the rounding 
of the optimal solution of a suitable linear program. In the case of objects of non-uniform
length, the authors proved that the problem of deciding whether an instance admits a solution is
NP-complete and provided a polynomial 20.5-approximation algorithm that produces a solution at 
which the capacity of each client's cache exceeds its capacity in the optimum solution by at most the length of the largest
object. The approximation ratio for unit-sized objects was later improved to 10 in~\cite{Swamy04} and~\cite{Baev08}. 

Various previous works have also considered variants of the data placement problem in terms of the 
underlying topology. In~\cite{Leff93} the authors consider the case of distances 
in the underlying topologies that form an ultrametric, i.e. are non-negative, symmetric and satisfy 
the {\em strong} triangle inequality, that is $d(i,j)\leq \max\{d(i,k), d(k,j)\}$ for clients $i,j,k$. 
The authors consider a simple hierarchical network consisting of three distances between the clients and 
devise a polynomial algorithm for the case of unit-sized objects by transforming it to a capacitated transportation
problem~\cite{Garfinkel73}. For the case of general ultrametrics, an optimal polynomial algorithm is 
given in~\cite{Korupolu01} based on a reduction to the min-cost flow problem. 

The {\it page-placement problem} is an important generalization of the data placement and was proposed 
and studied in~\cite{Meyerson01}. In this problem, each client has an extra constraint on the number of other 
clients it can serve, apart from the constraint on the capacity of its cache. 
In~\cite{Meyerson01}, the authors give $5$-approximation algorithm for the 
problem which violates both client and cache capacity constraints by a logarithmic factor at most. 
In~\cite{Guha02}, the logarithmic violation of both capacity constraints was improved to constant 
with a $13$-approximation algorithm. Finally, in~\cite{Laoutaris06} and~\cite{Pollatos08} a game-theoretic 
aspect of the data placement problem is studied, where clients are considered to be selfish agents. In both
works, algorithms are provided which stabilize clients in equilibrium placements. 

All previous results capture situations where write requests are rarely or never issued for
the objects. In~\cite{Baev08} the authors consider the case when write requests are common and  
formulate the {\em connected data placement} problem, in which it is required that all replicas of 
an object $o$ are connected via a Steiner tree $T_o$ to a root $r_o$, which can later be used as a
multicast tree. The objective is the minimization of the total incurred access cost and the cost 
of building the Steiner tree. A $14$-approximation algorithm for the problem is given in~\cite{Baev08}. 
This problem is a generalization of the {\em connected facility location problem} for which the best known 
approximation ratio is 8.55~\cite{Swamy04b}.
		
In section 2 we formally define the DP problem and introduce appropriate notation. 
In section 3 we present our main results for the DP problem, whereas in section 4 we present 
an algorithm for the page placement problem and also briefly discuss modifications for the other
extensions.

\section{Problem definition}

The data placement problem we consider in this paper is identical to the one in~\cite{Baev01} and
is abstracted as follows\footnote{In~\cite{Baev08} a seemingly different but essentially equivalent formulation 
of the problem is described.}. 
There is a network $\mathcal{N}$ consisting of a set $\mathcal{M}$ of 
$M=|\mathcal{M}|$ users (clients) and a universe $\mathcal{O}$ of $N=|\mathcal{O}|$ objects. In what follows 
we use the terms user and machine interchangeably. Each object $o\in\mathcal{O}$ has length $l_o$ and each user $j\in\mathcal{M}$
has a local capacity $C_j$ for the storage of objects. The distance between the users can be represented by a distance
matrix $D$ (not necessarily symmetric) where $d_{ij}$ denotes the distance from $j$ to $i$. The
matrix $D$ models the underlying topology. We do not assume any restrictions (e.g. metric) on the
distances. Each user $i$ requests access to a set of objects $R_i\subseteq \mathcal{O}$, namely
its {\em request set}. For each object $o$ in its request set, client $i$ has a {\em demand} of
access $w_{io}>0$. This demand can be interpreted as the frequency under which user $i$ requests
object $o$. The subset $P_i$ of its request set, that $i$ chooses to replicate locally is referred
to as its {\em placement}. Obviously, $|P_i|\leq C_i$ for unit-sized objects. We assume an
installation cost $f^o_i$ for each object $o$ and each cache $i$. The objective is to choose
placements of objects for every client such as the total induced access and installation costs for
all objects	and all clients is minimized. In the following, we will assume without loss of
generality that each	object $o\in {\mathcal O}$ is requested by at least one user. 
	
We define a {\em configuration} $c\subseteq {\mathcal M}$ as a
(non 	empty) subset of the $M$ machines. Thus, we have $2^M-1$ distinct configurations and we denote by 
$\mathcal{C}$ the set of all configurations. For a configuration $c\in\mathcal{C}$ and a user $j$
we say that $j$ {\em is used with respect to} $c$, denoted by $j\in c$, if the configuration $c$ 
contains $j$'s cache, i.e. machine $m_j\in c$. It will be also convenient to introduce the
following notation: $p_{cj}=1$ if $j\in c$, and $p_{cj}=0$ otherwise. For an object $o$, we define a 
{\em $c$-placement} with respect to $o$, as a placement of object $o$ to the machines belonging to 
$c$.

Introducing binary variables $x_{oc}$ to denote whether we choose or not the $c$-placement with 
respect to $o$, we can formulate our problem as an integer linear program, denoted by ILP in the 
sequel, in the following way:
\begin{equation}
\begin{array}{lcl}
\mbox{minimize}   & \multicolumn{1}{c}{\displaystyle \sum_{o\in\mathcal{O}}\sum_{c\in
\mathcal{C}}cost_{oc}\: x_{oc}} & \\
\mbox{subject to} & \displaystyle \sum_{o\in\mathcal{O}}\sum_{c\in \mathcal{C}}l_o\: p_{cj}\:
x_{oc}\leq C_j\ \  & j\in \mathcal{M}\\
& \displaystyle \sum_{c\in \mathcal{C}}x_{oc}=1 & o\in \mathcal{O}\\
& x_{oc}\in\{0,1\} & o\in \mathcal{O},\; c\in \mathcal{C}
\end{array}
\tag{ILP}
\end{equation}
where $cost_{oc}$ is the total cost induced if configuration $c$ is used for the placement of
object $o$, that is $cost_{oc}=\sum_{j\in\mathcal{M}}(1-p_{cj})\: w_{jo}\: l_o \: d_j(c) + 
\sum_{j\in \mathcal{M}} p_{cj} \: f^o_j$, with $d_j(c)= \min_{j'\: : \: p_{cj'}=1}d_{j'j}$ the 
nearest distance at which client $j$ can access object $o$ under configuration $c$. The first set
of constraints essentially states that the set of objects that each user replicates must not violate 
the user's cache constraint, while the second set states that for each object exactly one 
configuration should be chosen. In what follows we denote by $OPT$ the optimum solution of the 
previous program.

\comment{\begin{figure}[t]
\center
\psfrag{o1}{\fs{$o_1$}}
\psfrag{o2}{\fs{$o_2$}}
\psfrag{o3}{\fs{$o_3$}}
\psfrag{ok}{\fs{$o_k$}}
\psfrag{ok+1}{\fs{$o_{k+1}$}}
\psfrag{oN}{\fs{$o_N$}}
\psfrag{s}{\fs{$s$}}
\psfrag{t}{\fs{$t$}}
\psfrag{c1}{\fs{$c_1$}}
\psfrag{c2}{\fs{$c_2$}}
\psfrag{cm2}{\fs{$c_{2^M}$}}
\psfrag{cost1c1}{\fs{$cost_{o_1c_1}$}}
\psfrag{cost1cm}{\fs{$cost_{o_1c_{2^M}}$}}
\psfrag{cost2c1}{\fs{$cost_{o_2c_1}$}}
\psfrag{cost2cm}{\fs{$cost_{o_2c_{2^M}}$}}
\psfrag{costkc1}{\fs{$cost_{o_kc_1}$}}
\psfrag{costkcm}{\fs{$cost_{o_kc_{2^M}}$}}
\psfrag{costnc1}{\fs{$cost_{o_Nc_1}$}}
\psfrag{costncm}{\fs{$cost_{o_Nc_{2^M}}$}}
\psfrag{1s}{\fs{$0$}}
\includegraphics[scale=0.4]{dp.eps}
\caption{\fs{Representation of the data placement problem with constant number of clients as a 
constrained shortest path problem. Each edge connecting two nodes $c_io_k$ and $c_jo_{k+1}$ has a 
weight equal to the cost $cost_{o_kc_i}$.}} 
\label{fig:csp}
\end{figure}}

Note that the problem, as defined above does not always admit a feasible solution. In order to avoid trivial
cases of infeasibility we assume in the sequel that $\sum_{i\in\mathcal{M}}C_i\leq\sum_{o\in\mathcal{O}}l_o$
which essentially states that all clients can collectively store the union of the requested objects.
Other works (\cite{Leff93,Korupolu01}) assume existense of a distant server, that is, a user holding as a fixed 
placement the universe of objects, which essentially tackles the problem of infeasibility. For the case of
uniform sized objects, this assumption has no effect in the problem's hardness since the hardness result
of Baev et al.~\cite{Baev01} also holds in this case. However, in the case of non-uniform sized objects,  
their result does not hold immediately, since it relies on the fact that it is sometimes not possible to find 
any feasible solution. When a distant server exists, any instance always admits a 
feasible solution. Nevertheless, their proof of non-approximability can be adapted and thus the following result
can be obtained. Due to space limitations, we defer the details of the proof to the full version of this paper.

\begin{proposition}
For any polynomial time computable function $\alpha(N)$, the data placement problem with non uniform 
object lengths and without any augmentation in cache capacities, cannot be approximated within a factor of 
$\alpha(N)$, unless {\bf P}={\bf NP}.
\end{proposition}

The problem can also be stated as a constrained shortest path problem as follows: we 
introduce a node for each binary variable $x_{oc}$ and two nodes $s$ and $t$ and connect them as 
follows: for each $o_i$, $1\leq i\leq N-1$ we connect the node that represents $x_{o_ic}$ with
every node that represents $x_{o_{i+1}c}$ for all $c$. Furthermore we connect node $s$ with nodes 
$x_{o_1c}$ and node $t$ with nodes $x_{o_Nc}$ for all $c$. 
\comment{(see also Fig.~\ref{fig:csp}).} 
At each 
edge $(x_{o_ic},x_{o_{i+1}c'})$ we assign a weight equal to $cost_{o_ic}$ for $1\leq i\leq N-1$. 
Edges $(x_{o_Nc},t)$ have weight $cost_{o_Nc}$ and edges $(s,x_{o_1c})$ have a weight of $0$. The 
objective is to find the shortest path between nodes $s$ and $t$ while respecting cache capacity 
constraints on each node. These constraints are assigned to each node $x_{o_ic}$ by simply summing 
up for each client contained in configuration $c$ the current cache contents up to object $o_i$. 
This constrained shortest path problem can be solved using dynamic programming. It leads to the 
algorithm presented in the next section.

\section{Constant number of clients}
\label{section:const}
In this section we focus in the case where the number of clients in the network (i.e. users) is a 
constant. To the best of our knowledge these are the first results for this natural variation. 
We show that the data placement problem can be solved optimally in polynomial
time when all objects are unit-sized. When objects have different sizes we are still able to solve 
the problem optimally, with only a small and asymptotically tight violation of the cache
capacities. 

\subsection{Uniform length objects}
Let us define an available cache vector $r=(r_1,r_2,\ldots, r_M)$, where $r_j$ denotes the current 
space size available on cache of user $j$, for $1\leq j\leq M$. For $1\leq k\leq N$, let us denote 
by $f_k(r)$ the cost associated with the optimal way of placing objects $o_{1},\ldots,o_{k}$ on the 
clients' caches, assuming that the current available cache vector is $r$. For any configuration
$c$, we denote by $\delta_c=(\delta^1_c,\ldots, \delta^M_c)$ its machine-profile vector, with 
$\delta^i_c=1$ if configuration $c$ uses machine $m_i$, and $\delta^i_c=0$ otherwise. We assume in 
this section that all lengths satisfy $l_{o}=1$, but the following recurrence holds for the general 
case and it will be also used in the next section. One has
$$f_k(r)=\min_{c\: :\: r - l_{o_k} \delta_c \geq 0} (cost_{o_k c} + f_{k-1}(r- l_{o_k} \delta_{c})),$$
with $f_0(r)=0$ for any $r$. Finding the optimum cost to ILP reduces to the computation of $f_N(r)$ 
with $r=(C_1,C_2,\ldots C_M)$.

\begin{theorem}\label{thm:unit}\hfill
The non-metric data placement problem with uniform length objects and a fixed number of clients can 
be solved optimally in polynomial time.
\end{theorem}
\begin{proof}
By using standard techniques (see for example~\cite{kleinbergtardosbook}), the above recurrence 
leads to an efficient dynamic programming algorithm to obtain the optimal cost and solution of ILP.
The cache vectors $r$ can take values from a set of size $\prod_{j=1}^MC_j\leq C^M_{max}$ where 
$C_{max}$ is the maximum cache size. Assuming the values $f_k(r)$ are stored in an array and 
computed from $k=1$ to $k=N$, then for each $r$ the time needed to compute $f_k(r)$ is $O(2^M)$, 
i.e. a constant time, since at most $2^M$ configurations need to be checked. The total time 
complexity is therefore $O(N2^MC_{max}^M)$. Notice that since objects are unit-sized, i.e. $l_o=1$, 
$\forall o\in \mathcal{O}$, we can assume without loss of generality that for any capacity we have 
that $C_j \leq N$. If it is not the case, by changing this capacity to $C_j := N$, we obtain an 
equivalent instance because in the model considered, a client has no incentive to replicate any 
distinct object twice, since this would have no effect in the total access cost. Finally, the 
computation time becomes $O(N^{M+1})$.\qed
\end{proof}

\subsection{Non-uniform length objects} \label{nusect}
The previous dynamic programming algorithm is in fact pseudo-polynomial, since the complexity 
$O(N^MC_{max}^M)$ depends on the maximum cache size $C_{max}$. In the case of unit-sized objects we 
are able to bound $C_{max}$ by the total number of objects and thus obtain a polynomial time 
algorithm. In the case of objects of arbitrary length the bound $C_{max}\leq N$ does not hold and 
the algorithm remains pseudo-polynomial.

\begin{algorithm}[t]
$\alpha\gets(\varepsilon l_{max})/N$\;
\tcp{update object lengths}
\ForEach{$o\in \mathcal{O}$}{$l'_o\gets\left\lfloor l_o/\alpha\right\rfloor$\;}
\tcp{update cache sizes}
\ForEach{$j\in \mathcal{M}$}{$C'_j\gets\left\lfloor C_j\alpha\right\rfloor$\;}
\tcp{use updated lengths and cache sizes with dynamic programming}
$OPT_{\alpha}\gets$ optimum solution of $\mbox{ILP}_{\alpha}$\;
Output $OPT_\alpha$\;
\caption{DP-NU($\mathcal{M},\mathcal{O},\varepsilon$)} 
\label{alg:dpnu}
\end{algorithm}

In what follows, we show how to design a polynomial time algorithm in the case of arbitrary-sized 
objects.  We let $\alpha = \varepsilon l_{max}/N$ where $\varepsilon$ is an arbitrarily small 
positive constant and modify the object lengths and cache sizes appropriately. To compute a
solution we use algorithm~\ref{alg:dpnu} where $\mbox{ILP}_\alpha$ denotes the integer linear program obtained 
from ILP by using length $l'_o$ (resp. cache $C'_j$) instead of $l_o$ (resp. $C_j$) for all objects 
$o$ and clients $j$. 

Notice however that the cost function in $\mbox{ILP}_\alpha$ is the same as in 
ILP, i.e. the costs $cost_{oc}=\sum_{j\in\mathcal{M}}(1-p_{cj})\: w_{jo}\: l_o \: d_j(c) 
+ \sum_{j\in \mathcal{M}} p_{cj} \: f^o_j$ are calculated by using the initial lengths $l_o$. We 
have the following lemma.
\begin{lemma}\label{lemma1}
Given an $\alpha>0$, any solution $x$ for ILP is a solution for $\mbox{ILP}_{\alpha}$.
\end{lemma}
\begin{proof}
Let $x$ be a solution of ILP. One has, $\forall j\in \mathcal{M}$, 
\begin{equation}
\sum_{o\in\mathcal{O}} \sum_{c\in \mathcal{C}} \left\lfloor 
\frac{l_o}{\alpha}\right\rfloor \: p_{cj}\: x_{oc} \leq 
\sum_{o\in\mathcal{O}} \sum_{c\in \mathcal{C}} \left\lfloor 
\frac{l_o}{\alpha} \: p_{cj}\: x_{oc}\right\rfloor \leq 
\left\lfloor \sum_{o\in\mathcal{O}} \sum_{c\in \mathcal{C}} 
\frac{l_o}{\alpha} \: p_{cj}\: x_{oc}\right\rfloor \leq 
\left\lfloor \frac{C_j}{\alpha}\right\rfloor, \label{eqlem1}
\end{equation}
where the first inequality comes from the fact that $p_{cj}$ and $x_{oc}$ are integers, the second 
inequality is a standard one, and the last inequality comes from $\sum_{o\in\mathcal{O}} \sum_{c\in 
\mathcal{C}} l_o\: p_{cj}\: x_{oc}\leq C_j$ since $x$ is a feasible solution of ILP. Therefore, $x$ 
satisfies $\sum_{o\in\mathcal{O}} \sum_{c\in \mathcal{C}} l'_o\: p_{cj}\: x_{oc}\leq C'_j$, and $x$ 
is a feasible solution of $\mbox{ILP}_\alpha$.\qed
\end{proof}

From the above lemma, we can immediately conclude that if $\mbox{ILP}_\alpha$ has no solutions,
then the same holds for ILP. However, if ILP has no feasible solutions, $\mbox{ILP}_\alpha$ could have 
feasible solutions. In the following, we assume that ILP admits at least one feasible solution, in 
order to be able to define an optimal solution denoted by OPT.

\begin{lemma}\label{lemma2}
The algorithm DP-NU$({\mathcal M},{\mathcal O})$ returns an optimal solution for ILP using 
$\varepsilon l_{max}$ blow-up in time polynomial in $N$ and $1/\varepsilon$, where $\varepsilon$ is 
an arbitrarily small positive constant and $l_{max}$ is the length of the largest object.
\end{lemma}
\begin{proof}
First, notice that by Lemma~\ref{lemma1} the cost of the solution $\mbox{OPT}_\alpha$ is not
greater 
than the cost of the solution OPT. Furthermore, we have that
\begin{eqnarray*}
\sum_{o\in\mathcal{O}}\sum_{c\in \mathcal{C}}\left\lfloor\frac{l_o}{\alpha}\right\rfloor p_{cj}
\: x_{oc}\geq\sum_{o\in\mathcal{O}}\sum_{c\in \mathcal{C}}\left(\frac{l_o}{\alpha}-1\right)\:
p_{cj}\: x_{oc}
\geq \sum_{o\in\mathcal{O}}\sum_{c\in \mathcal{C}}\frac{l_o}{\alpha}\: p_{cj}\: x_{oc}-N
\end{eqnarray*}
which becomes
\begin{eqnarray*}
\sum_{o\in\mathcal{O}}\sum_{c\in \mathcal{C}}l_o\: p_{cj}\: x_{oc}&\leq&\alpha\:\sum_{o\in
\mathcal{O}}\sum_{c\in \mathcal{C}}\left\lfloor\frac{l_o}{\alpha}\right\rfloor\: p_{cj}\: 
x_{oc}+\alpha\: N\\
&\leq& \alpha\:\left\lfloor\frac{C_j}{\alpha}\right\rfloor+N\:\alpha\leq C_j+N\:\alpha \ \mbox{\ \ \
(by using inequality~(\ref{eqlem1}))}
\end{eqnarray*}
Putting $\alpha = \varepsilon l_{max}/N$ we get for the initial instance
\begin{equation}
\sum_{o\in\mathcal{O}}\sum_{c\in \mathcal{C}}l_o\: p_{cj}\: x_{oc}\leq
C_j+ \varepsilon l_{max}
\label{eq:blow}
\end{equation}
thus, each cache size is violated by at most $\varepsilon l_{max}$.

For the complexity, notice that for any user $j$'s cache, we can assume without loss of generality 
that $C_j \leq N l_{max}$. If it is not the case, by changing the capacity to $C_j := N l_{max}$,
we obtain an equivalent instance because in the model considered, a client has no incentive to
replicate any distinct object twice, since this would have no effect in the total access cost.
We have therefore, 
$$ C'_j = \left\lfloor\frac{C_{j}}{\alpha}\right\rfloor\leq 
\frac{C_j}{\alpha} \leq\frac{N l_{max}}{\alpha}\leq \frac{N^2}{\varepsilon}.$$
Finally, we obtain $C'_{max} = \max_{j\in {\mathcal M}} C'_j \leq N^2/\varepsilon$ and by a similar 
analysis as in theorem~\ref{thm:unit}, the complexity of $O(N2^MC_{max}^M)$ becomes 
$O(N^{2M+1}\varepsilon^{-M})$. Notice that if $\alpha$ is large enough, some lengths $\lfloor
l_o/\alpha\rfloor$ can become equal to zero. In that case, the dynamic programming algorithm can be 
accelerated for such objects, since an optimal placement is to put them on each machine.\qed
\end{proof}

\noindent Using Lemma~\ref{lemma2} we obtain the following theorem.

\begin{theorem}
The non-metric data placement problem, with non-uniform object lengths and a fixed number of 
clients, can be solved optimally in polynomial time using $\varepsilon l_{max}$ blow-up on the 
machines' capacity, where $\varepsilon$ is an arbitrarily small positive constant and $l_{max}$ is 
the length of the largest object.
\end{theorem}

The $\varepsilon l_{max}$ blow-up stated in the previous theorem is asymptotically tight. In order
to clarify this, consider an instance with $N$ objects and two clients $M_1$ and $M_2$.
The lengths of the objects are $l_i=(1-\delta)/N$ for $i=1,\ldots,N-1$ and $l_N=l_{\max}=1/\epsilon$ 
where $0<\epsilon<1$ and $0<\delta<1$. The cache capacities of the clients 
are $C_1=\epsilon l_{\max}=1$ for $M_1$ and $C_2=1/\epsilon$ for $M_2$. All installation costs are 0. Client $M_1$ has a demand 
equal to 1 for the first $N-1$ objects and no demand for object $N$. Client $M_2$ 
has also a demand of 1 for the first $N-1$ objects and a demand of $N$ for the $N$-th object. 
In the optimum solution $OPT$, $M_1$ replicates all the $(N-1)$ objects and $M_2$ replicates only
object $N$. When our algorithm is employed, the lengths of objects $i$, $1\leq i\leq N-1$ 
become $l'_i=\left\lfloor l_i/\alpha\right\rfloor=
\left\lfloor 1-\delta\right\rfloor=0$. The length of the $N$-th object becomes
$l'_{\max}=\left\lfloor N/\epsilon\right\rfloor$ while the cache sizes become
$C'_1=\left\lfloor 1/\alpha\right\rfloor=N$ and $C'_2=
\left\lfloor N/\epsilon\right\rfloor$. In the optimum solution $OPT_\alpha$ client 
$M_1$ will again choose to replicate the $(N-1)$ objects it has demand for, but 
client $M_2$ can now choose all $N$ objects. After restoring the original object lengths
and capacities, the total blow-up is only due to $M_2$ and is equal to
$(N-1)((1-\delta)/N)$. Choosing $\delta=1/(N-1)$ we get $1-2/N$
the limit of which is $1=\epsilon\cdot\frac{1}{\epsilon}=\epsilon\cdot l_{\max}$, as $N$ 
approaches infinity.

\section{The page placement problem}
In the page placement problem, there are bounds imposed on the number of clients that can connect to 
a specified client's cache in order to access objects. We denote by $k_j$ the maximum number of users 
that can access a given user $j$'s cache. If the same user access cache $j$ for different objects it 
is counted only once. Clearly, in this problem a client requesting an object can not always use the 
nearest machine which replicates that object to access it.

We need to introduce some terminology and notations. Let us define an available load vector 
$t=(t_1,t_2,\ldots, t_M)$, where $t_j$ denotes $k_j$ minus the current number of users connected to 
the cache $j$. Notice that the number of load vectors is bounded by $\prod_{j=1}^M (k_j+1)\leq 
(M+1)^M$. For any configuration $c$, we denote as before by $\delta_c$ its machine-profile vector, 
i.e. for $1\leq i\leq M$, $\delta^i_c=1$ if configuration $c$ uses machine $m_i$, and $\delta^i_c=0$ 
otherwise.

Given an object $o$ and a configuration $c$, a $c$-placement is a placement such that a machine $m$ 
receives the object $o$ if and only if $m\in c$. In a $c$-placement, the machines outside $c$ need a 
way to access the object $o$ they are requesting. We call such a way a {\it connection pattern} 
$\rho$ with respect to the configuration $c$, and we denote by $\Phi_c$ the set of all such possible 
connection patterns. Given $\rho\in \Phi_c$, for all $j\notin c$ and $i\in c$, we put $\rho_{ij}=1$ 
if user $j$ access object $o$ from user $i$, and $\rho_{ij}=0$ otherwise. Moreover, for all $j\in c$ 
and $i$ we have $\rho_{ij}=0$. Notice that $|\Phi_c|$ is bounded by $|c|^{M-|c|} \leq M^M$. Finally, 
we define a {\em history pattern} $s$ in the following way: $s_{ij}=1$ if machine $m_j$ has 
previously used machine $m_i$ to access an object, and $s_{ij}=0$ otherwise. The number 
of history patterns is equal to $2^{M(M-1)/2}$. Given $\rho$ and $s$ we denote by $s\vee \rho$ the 
updated history pattern taking into account the current connection pattern $\rho$. The updated pattern 
can be obtained by performing a logical {\it or} between $\rho$ and $s$, i.e.  
$(s\vee \rho)_{ij} = s_{ij}\vee \rho_{ij}$.

For a connection pattern $\rho\in \rho_c$ and a history pattern $s$, we denote by 
$\Delta_{\rho,s}=(\Delta_{\rho,s}^i)_{i=1}^M$ the vector which indicates for each machine $m_i$ the 
number of machines which are connected to $m_i$ for the first time. Such a vector can be obtained in 
the following way: for $1\leq i\leq M$, one has $\Delta_{\rho,s}^i = \sum_{j=1}^M \rho_{ij} 
(1-s_{ij})$. Finally, we define $cost_{o,c,\rho} = \sum_{i,j\in {\mathcal M}} d_{ij} w_{jo} l_o \rho_{ij} + 
\sum_{i\in c} f_i^o$.

For $1\leq k\leq N$, let us denote by $f_k(r,t,s)$ the cost associated with the optimal way of 
placing objects $o_{1},\ldots,o_{k}$ on the clients' caches, assuming that the current available 
cache vector, load vector and access vector are respectively $r,t,s$. One has
$$f_k(r,t,s)=\min_{\substack{C\in {\mathcal C} :\\r - l_k \delta_c \geq 0}} \ \ 
\min_{\substack{\rho\in \rho_c :\\t-\Delta_{\rho,s}\geq 0}} \ \left(cost_{o_k,c,\rho} +
f_{k-1}(r-l_{o_k} \delta_{c},\, t-\Delta_{\rho,s},\, s\vee \rho)\right),$$
with $f_0(r,t,s)=0$ for any $r,t,s$.

\begin{theorem}
The non-metric page placement problem with uniform length objects and a fixed number of clients
can be solved optimally in polynomial time.
\end{theorem}
\begin{proof}
Finding the optimum cost to the problem reduces to the computation of $f_N(r,t,s)$ with $r=(C_1,\ldots C_M)$,
$t=(k_1,\ldots, k_M)$ and $s=(0,\ldots 0)$. The complexity for computing $f_k(r,s,t)$ is $O(2^M\, M^M\, M)$,
and there are at most $N C_{max}^M M^M 2^{M(M-1)/2}$ $(r,s,t)$ triplets. As in 
section~\ref{section:const} (Theorem~\ref{thm:unit}), we can assume that $C_{max}\leq N$ and obtain 
an overall complexity of $O(N^{M+1})$.\qed
\end{proof}
For the non uniform case, the same recurrence relation holds, and using a similar technique and analysis as in 
section~\ref{nusect} (not repeated here due to space limitations) the complexity becomes $O(N^{2M+1} \varepsilon^{-M})$ 
and we obtain the following result:

\begin{theorem}
The non-metric page placement problem with non-uniform object lengths and a fixed number of clients, can
be solved optimally in polynomial time using $\varepsilon l_{max}$ blow-up on the 
machines' capacity, where $\varepsilon$ is an arbitrarily small positive constant and $l_{max}$ is 
the length of the largest object.
\end{theorem}

\comment{\noindent {\em 4.1 Existence of a distant server.}\\
In this variant, we assume existense of a machine (server) whose cache's capacity is arbitrarily 
large and can therefore always hold all the objects. In~\cite{Baev08} a non approximability result 
is provided for the DP placement with non uniform 
object lengths. It does not hold immediately for this case, since it relies on the fact that it is
sometimes not possible to find any feasible solution. In our case, any instance always admits a 
feasible solution because of the existence of the server machine. However, the proof of non 
approximability can be easily adapted. We detail the proof in Appendix A.
\begin{proposition}
For any polynomial time computable function $\alpha(N)$, the data placement problem with non uniform 
object lengths, and without any blow-up in cache, cannot be approximated within a factor of 
$\alpha(N)$, unless {\bf P}={\bf NP}.
\end{proposition}}
 
\section{Concluding Remarks}
In this paper, we addressed the problem of replicating data over a constant number of network clients and 
designed optimal algorithms via utilization of the notion of configurations. If 
all data objects are equal in size, our algorithm finds in polynomial time the optimum solution. When lengths 
of objects differ, a small violation of each client's cache capacity constraint is enough, so as to be able to find the 
optimum solution. 

Our technique constitutes a general framework that can also be used for solving optimally various common 
extensions of the problem such as: (a) the $k$-median variant in which 
an upper bound $k_o$ is imposed on the number of copies of each object $o$ that can be replicated in the network and
(b) the connected data placement problem~\cite{Baev08}, where apart from placing objects, all clients 
holding replicas of the same object should also be innterconnected via a directed Steiner tree. 
Furthermore our technique can be applied for other variants of data placement for example the fault 
tolerant data placement (derived from the fault-tolerant facility location problem~\cite{ft2}) 
where each client can be served by a given number of machines and the cost is obtained by summing 
the costs of access with respect to those machines. We defer the details for 
these and other extensions, due to space limitations, for the full version of this paper.
 
The proposed algorithms remain polynomial independently of any metric. An important aspect of further research 
is the modification of the described algorithm so as to be able to handle extensions involving payments. In such 
extensions, apart from object preferences, a client also has a budget to spend and pay other 
clients to convince them to replicate certain objects.

\end{document}